# Tracing Cryptocurrency Scams: Clustering Replicated Advance-Fee and Phishing Websites

Ross Phillips and Heidi Wilder
*Elliptic*
London, UK
{r.phillips, h.wilder}@elliptic.co

*Abstract*—Over the past few years, there has been a growth in activity, public knowledge, and awareness of cryptocurrencies and related blockchain technology. As the industry has grown, there has also been an increase in scams looking to steal unsuspecting individuals' cryptocurrency. Many of the scams operate on visually similar but seemingly unconnected websites, advertised by malicious social media accounts, which either attempt an advance-fee scam or operate as phishing websites. This paper analyses public online and blockchain-based data to provide a deeper understanding of these cryptocurrency scams. The clustering technique DBSCAN is applied to the content of scam websites to discover a typology of advance-fee and phishing scams. It is found that the same entities are running multiple instances of similar scams, revealed by their online infrastructure and blockchain activity. The entities also manufacture public blockchain activity to create the appearance that their scams are genuine. Through source and destination of funds analysis, it is observed that victims usually send funds from fiat-accepting exchanges. The entities running these scams cash-out or launder their proceeds using a variety of avenues including exchanges, gambling sites, and mixers.

## I. Introduction

The past few years have seen increasing awareness and participation in the blockchain and cryptocurrency industry, especially from previously unacquainted audiences. The total number of user accounts at cryptocurrency service providers was estimated to exceed 139 million in late 2018, with individuals (as opposed to businesses) still constituting the largest share of their customers [1]. Accompanying this increase in individuals holding, using, and speculating with cryptocurrencies, there has also been an increase in the number of malicious entities aiming to extract cryptocurrency funds (henceforth referred to as funds) from unsuspecting victims. Techniques employed by these entities include hacking exchanges or individuals [2], ransomware [3], and running Ponzi schemes [4].

One way entities may attempt to extract cryptocurrency funds from victims is through the use of scams. Thousands of seemingly isolated scam websites follow one of two general strategies: (1) The website attempts to convince a victim to transfer cryptocurrency to a blockchain address with the promise of returning more cryptocurrency. This structure is referred to widely as an *advance-fee* scam or, specific to the cryptocurrency ecosystem, a *trust trading* scam. (2) The website convinces a victim to provide existing credentials required to access their cryptocurrency with the promise of funds or access to a service. Phishing websites follow this second format; the victim believes they are on a legitimate website but are instead revealing their credentials to a malicious entity.

Unsuspecting victims are usually directed to these scam websites via links on social media. It is common to observe well-known individuals such as Elon Musk, John McAfee, and Vitalik Buterin on Twitter purportedly offering free cryptocurrency. However, on closer inspection, imposters run the accounts. One study tracked the evolution of over 15,000 Twitter accounts set up to propagate such cryptocurrency scams [5]. The prevalence of submissions advertising these scams has led Twitter to release guidelines addressing financial scams in September 2019.[1]

This paper aims to provide an understanding of these specific cryptocurrency scams by addressing the following questions: (1) Can the plethora of isolated scam websites be broken down into distinct types? (2) Are scam websites standalone or part of more extensive connected campaigns operated by the same entity or entities? (3) Where are scammed funds being sent from? (4) Where are scammed funds being sent to?

To answer these questions, Open Source Intelligence (OSINT) tools are used to extract information about the scam websites. A clustering technique is first applied to the content of scam websites to group scams into different types. The same clustering technique is then applied to registration and ownership details of the websites to identify similarities in their ownership data. Further information about the funds received are analysed through blockchain analysis, which is made possible due to the transparent nature of the blockchains involved and would be unachievable if fiat currencies were the currency of choice for the scam websites.

The remainder of this paper is structured as follows: Section II reviews the background and relevant literature on cryptocurrency-specific scams and how they have been previously analysed through online and blockchain-based techniques. Section III explores the data sources used and the data extraction process. Section IV examines the DBSCAN clustering methodology used and details the considerations made while applying it. Section V presents the results of the clustering algorithm as well as the further analysis done. Section VI concludes with a discussion.

## II. Background

This section details illicit cryptocurrency activity within the broader context of illicit online activity, before considering tracing cryptocurrency-specific activity using blockchain analytics.

---

[1] https://help.twitter.com/en/rules-and-policies/financial-scam

*A. Illicit online activity including cryptocurrency activity*

Illicit online activity, and more generally cybercrime, continuously evolves to take advantage of changing technologies [6]. Such activity can be split into three separate types [7]: 1) Traditional forms of crime conducted over the internet; 2) publication of illegal content distributed over the internet; 3) crimes unique to the internet, i.e. hacks and botnets.

As with the advent of the internet, the advent of cryptocurrencies has provided opportunities for existing crimes to evolve and find a new platform, payment system or niche. These crimes form a small minority of the use of cryptocurrencies compared to the plethora of legitimate and positive impacts the technology can have (one estimate, on a subset of activity, is that 2% of cryptocurrency transactions are known to be illicit [8]).

One example where existing illicit activity has transitioned to use cryptocurrencies is ransomware, which since adopting cryptocurrencies have seen increased revenue [6]. Another example would be Ponzi schemes, which exist outside cryptocurrencies, but have managed to raise large amounts using cryptocurrencies too; for example, a large cryptocurrency-based Ponzi scheme called PlusToken attracted over $2 billion of investment. The advance-fee scams focussed on in this work first originated in the 1970s, where they were originally distributed via letter or fax before increasing in prevalence due to the growth of the internet [9] and some now involve cryptocurrencies. Pump and dump events are another example as they are seen occasionally in traditional financial markets. The price of a cryptocurrency is increased by a coordinated and sudden increase in demand. Those orchestrating such schemes announce the certain cryptocurrency and participants buy with the hope that they can sell at a higher price [10]. Participants are usually made aware of such schemes through invite links, distributed on mass by automated social media accounts [11].

As well as illicit activity involving existing cryptocurrencies, the increase in investment into Initial Coin Offerings (ICOs) seen in 2017 and 2018, provided some with the opportunity to invent their own cryptocurrency project, raise investment via existing cryptocurrency and not deliver on their promises [12], which could be considered similar to other non-cryptocurrency types of investment scams.

One form of illicit activity that appears unique to cryptocurrencies and decentralised blockchain applications is the dissemination of deceptive smart contracts. For example, smart contracts have been implemented to appear like fair Ponzi or pyramid schemes, these claim to be trustworthy as participants can usually view and verify their immutable logic, however participants still commonly lose their money [13]. Another example is the creation and dissemination of honeypot smart contracts which appear to contain vulnerabilities, however any funds sent to the contract as an attempt to exploit them are retained due to hidden traps [14].

*B. Blockchain analysis of illicit cryptocurrency flows*

Cryptocurrency addresses can be associated with illicit activity by extracting information about the addresses from online sources, including directly from illicit websites advertising the addresses or forum posts. For example, one study extracted addresses from posts pertaining to gambling, high-yield investment programs, and scam accusations on the cryptocurrency related forum bitcointalk.org [4]. Sometimes the original websites advertising cryptocurrency addresses may no longer be functional. In one such case, the Internet Archive is used to recover snapshots of Ponzi websites [15].

Given the public nature of the Bitcoin and Ethereum blockchains, once an address of interest has been identified, the magnitude of funds received can be examined [16]. In contexts where there are victims involved (for example, scams and ransomware), it is possible to estimate the number of suspected victims and the average amount lost by each victim [17]. Further analysis of the victim's payments can be done, for example by analysing the time of day and day of the week payments are made [3, 18]. As well as analysing the activity of victims, similar analysis can be applied to the perpetrator's activity. This is accomplished by examining outgoing transactions from an address of interest. Other schemes advertising addresses owned by the perpetrator can be searched for, as it has been found that addresses are reused between schemes [19]. New addresses may be generated for each victim, which makes it slightly harder to calculate the magnitude of all funds received immediately. Even when an entity is operating separate addresses, it is possible to group addresses suspected of being owned by the same entity through the use of rule-based heuristics [20] and machine learning [8]. It is also possible to track where funds are consolidated as funds in separate addresses may move to a single location; for example, consolidation addresses are used by ransomware operators [3]. The destination of where an entity sends its funds can also be examined. For example, previous work identified the types of services that funds predominantly flowing out of dark markets reached [21].

## III. DATA SOURCES

This section outlines the sources of data, used later in the clustering and blockchain analysis.

*A. Cryptocurrency scam collection and website analysis*

An initial list of cryptocurrency scam websites is sourced from CryptoScamDB. [2] This is a public repository of cryptocurrency-related scam reports, which is chosen as the starting source of data for replicability. Most of the scam websites listed on CryptoScamDB are no longer functional. This limited lifespan is observed elsewhere for websites used for illicit purposes [22]. As the websites no longer exist, further information is sourced from a free service called URLScan.[3] URLScan produces detailed snapshots of websites containing the website's raw HTML and visible page content, IP address, and outbound links. For each scam, URLScan is queried programmatically to check whether a relevant historical snapshot exists from when the scam was active. The details of the 4,139 relevant snapshots found are saved for use in the clustering; the exact features used are discussed in the methodology section (Section IV). URLScan also monitors and reports similar websites; 2,861 further scam websites are identified, bringing the total number of websites considered to 7,000. Fig. 1 shows the reported geolocation of the considered websites, it should be stressed that this indicates the location of the hosting provider, or the content delivery network (CDN), rather than the location of the scam operators.

---

[2] CryptoScamDB.org was previously called EtherScamDB and is occasionally referenced in academic literature as such. It rebranded in 2019. It is run by the MyCrypto team for the benefit of the cryptocurrency community. The raw data that this analysis uses is hosted on their GitHub, found here: bit.ly/38PButR

[3] URLScan.io

Historical domain registration data, relating to who registered the domain name of the website, are also collected. It should be noted that the dataset does not include every low-touch cryptocurrency scam website that exists, but is large enough to produce valuable findings.

### B. Blockchain analysis

Blockchain addresses are extracted from the snapshots of the scam websites via regular expression pattern matching. 18.1% of the websites have an extractable Bitcoin address advertised on their landing page and 38.7% of the websites have an extractable Ethereum address on their landing page. Each discovered address can potentially be associated with further addresses owned by the same entity using blockchain-based heuristics; for example, the common-spend heuristic and change heuristic [20]. We refer to addresses associated together in such a manner as a blockchain cluster, as cluster is the industry standard name; this is not related to a DBSCAN cluster.

Blockchain cluster information is sourced from Elliptic's dataset, which is built by combining forensic analysis of different blockchains with a growing proprietary dataset of addresses associated with known entities[4]. This knowledge of entities is important for the source of funds and destination of funds analysis, because it is used to identify those who are sending cryptocurrency to and receiving cryptocurrency from scams. Rather than naming entities individually, they are grouped into broader categories (for example, cryptocurrency exchange, dark markets, etc.). Exchanges are partitioned into fiat-accepting exchanges, cryptocurrency exchanges, or decentralised exchanges. This distinction is made as fiat-accepting exchanges are likely to have higher know your customer and anti-money laundering controls, whereas cryptocurrency exchanges and decentralised exchanges are likely to have more sophisticated users.

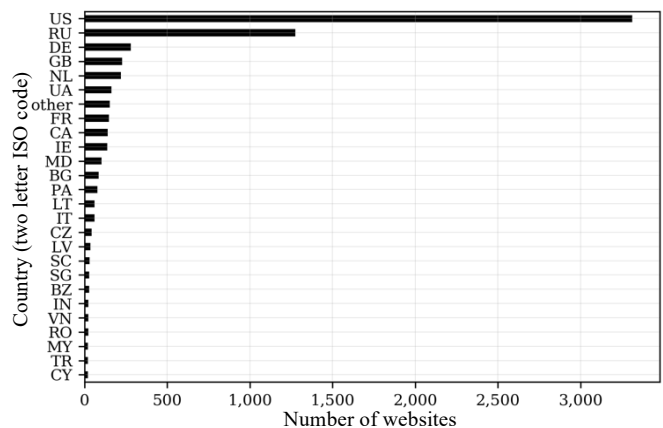

Fig. 1. Country geolocation of recorded website IP addresses (sourced from URLScan.io)

---

[4] For those wishing to conduct similar analysis, Bitcoin addresses grouped by cluster can also be retrieved from the free tool provided by oxt.me which also has certain labels (i.e. known entities) associated with particular blockchain addresses/clusters. Labels associated with certain Ethereum addresses are available at etherscan.io.

## IV. METHODOLOGY

This section outlines the chosen clustering technique, the considerations made while applying it, and details the exact features used.

### A. DBSCAN

The clustering algorithm chosen is DBSCAN [23], which, when applied to an n-dimensional dataset, identifies clusters as areas with a relatively higher density of data points. The DBSCAN algorithm is chosen because: (1) The number of clusters does not need to be specified *a priori*, which is the case in many other clustering algorithms; (2) the method does not require all points fall within clusters (data points in lower-density areas are labelled as outliers). This is likely to represent the selected population well as individual websites may exist that are not part of larger types or campaigns.

For an area of points to be considered a cluster, data points must fall within an epsilon (*Eps*) search radius distance from one another, and the area must contain a minimum number of points (*MinPts*). *Eps* and *MinPts* are two parameters that can be set and affect the number of identified clusters and the similarity of the points within the cluster. A suitable *Eps* is chosen through the method presented and visualised by [24]; the radius to nearest neighbours (RNN) is calculated for each data point, these values are then sorted, plotted, and the value at the point of maximum curvature is chosen. *MinPts* is left as the default value (5); we do not consider configuring this value as we plan to focus on the larger clusters and not those near this lower threshold.

### B. Feature preprocessing for scam type clustering

For clustering into scam types, we use the content of the scam websites. We generate our corpus from the URLScan content snapshots that provide the content already extracted from the raw HTML. Several pre-processing steps are applied to the content, including removing punctuation and stop words (for example, commonly used words such as "the"). Numbers are also removed, since it was determined that many of the scam websites use countdown timers and varying dates based on when the website is accessed (this causes arbitrary differences if included). Cryptocurrency addresses are also removed as they may be different even if two scams follow the same format. The standard TF-IDF algorithm [25] is then applied to get the final vectors for clustering.

### C. Feature preprocessing for campaign clustering

For clustering into campaigns, a variety of registration details and ownership identifiers are used, and are selected because they have been shown in previous work as suitable for clustering illicit websites. The selected attributes, outlined below, have been used individually before, and achieved successful results when grouping websites in a non-cryptocurrency specific context (for example [26, 27, 28]); this is the first known time they have been applied together and in the cryptocurrency domain.

*Google Analytics ID* - Many website owners use third-party analytics services to better understand how their website is used. These analytics services often require a user account ID to be placed within the source code of each of the user's websites they wish to monitor. Finding matching analytics service account IDs in the source code of seemingly unrelated websites indicate that they are reporting to the same analytics account. This has been shown recently as a way to cluster separate illicit websites into campaigns [26]. 17.7% of the websites in the dataset have a Google Analytics ID.

*Domain registrant details* - As done elsewhere [27], the email address is decomposed into two parts: the account (for example, the part of an email address before the @ sign) and the provider. This is done as both parts can separately provide information; for example, one campaign could use a different newly registered email account for each of its separate websites, but all from the same provider. The country of residence provided by the domain registrant is also used.

*Domain registrar* - This is the company via which domain names can be reserved and configured to map to IP addresses. A campaign being run by the same entity will likely exhibit patterns related to the use of registrars (e.g. repeatedly use the same registrar or switching between a couple of registrars [27]).

*IP address* - numerous separate domain names can be served content by the same IP address, and they are commonly used as a way to link illicit websites [28]. Using IP address as a feature allows investigation of whether the websites are being hosted from a single location or similar locations. The use of IP addresses as a feature within the clustering is likely to create higher confidence but smaller clusters.

The above features, some of which are text-based, need to be transformed into a numerical representation before clustering. Label encoding is applied to transform categorical text into numerical values; however, it cannot be applied alone as the clustering algorithm may misunderstand the new numerical labels to be following a meaningful order. For example, if registrar ABC and registrar XYZ get assigned to label 1 and label 2 respectively, the newly introduced order of the two labels and also their closeness may be used erroneously without further pre-processing. To overcome this, One Hot Encoding is also applied.

## V. RESULTS

This section first examines whether different scam types emerge when clustering based on their content. We then explore whether websites following the advance-fee format are part of the same organised campaigns. Once the scam websites have been clustered into types and separately into organised campaigns, blockchain forensics is used to analyse the proceeds they have raised, their evolution over time, and to find further connections between the websites.

### A. Clustering into types of scams

When clustering based on the website's content, 171 clusters are identified, representing different types or variants of the scam websites. On average, there are 24 websites per cluster; however, some significantly larger clusters emerge, which are focussed on here. Table I shows the 10 scams types with the largest number of occurrences. The left-most column contains the cluster number assigned by DBSCAN.

The largest type identified (type #25) contains 1,359 advance-fee websites; each website of this type (shown in Fig. 2) follows the same structure providing a payment address, QR code, and fake transaction logs purporting to be users successfully receiving funds. A relatively smaller type, type #39, is a modified version of type #25.

TABLE I. VARIANTS OF SCAMS IDENTIFIED

| # | Description | Members |
|---|---|---|
| 25 | Celebrity/Exchange Giveaway (Advance-Fee) | 1,359 |
| 0 | MyEtherWallet Clone - Pre-2019 redesign with pop-up (Phishing) | 352 |
| 1 | MyEtherWallet - Unclear Format (Phishing) | 173 |
| 51 | IDEX Clone (Phishing) | 101 |
| 5 | MyEtherWallet Clone - Pre-2019 redesign without pop-up (Phishing) | 80 |
| 93 | Coinbase Clone (Phishing) | 63 |
| 134 | Exchange Giveaway - Bespoke Format (Advance-Fee) | 60 |
| 100 | Exchange Giveaway – Blog Post (Advance-Fee) | 59 |
| 53 | Celebrity Giveaway – Blog Post (Advance-Fee) | 45 |
| 39 | Celebrity/Exchange Giveaway w/ Instructions (Advance-Fee) | 37 |

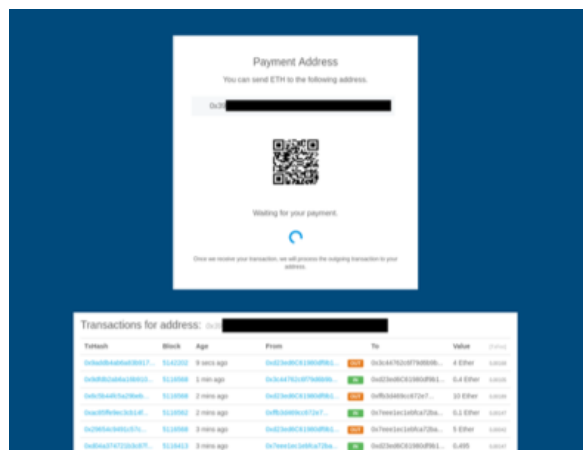

Fig. 2. The typical structure of a type #25 website

Type #39 uses the structure of type #25 but also includes instructions on how much cryptocurrency to send at the top of the webpages. Both type #25 and type #39 embed the gif of a loading spinner into their pages via hotlinking; this is sourced from the same location, highlighting that the websites are using the same resources. The gif is hosted on a generic image hosting site, where the provided metrics report that it has been requested over 2,560,000 times, giving an estimate for the number of impressions received by these particular websites. We investigate whether other websites are embedding this gif, finding a possible 10,000 scam websites, and in the process demonstrating a new way to discover further instances of these scams.

Type #53 and type #100 contain websites that appear like blog posts and advertise advance-fee scams purportedly from celebrities and exchanges, respectively. Another observed advance-fee cluster is type #134, which is similar to type #25, but uses bespoke websites with the logo and branding associated with a range of targeted exchanges.

Several types represent phishing pages targeting MyEtherWallet, a popular Ethereum-based web wallet. The separation into different types represent different variants; for example, the 2018 MyEtherWallet site is replicated with a pop-up window (type #0) and without (type #5). Types replicating exchange websites in an attempt to steal their customers credentials, including Coinbase (type #93), are also present.

## B. Clustering into organised campaigns

The campaign-based clustering is applied to the 1,560 websites identified in the previous section as running different types of advance-fee scams (i.e. the websites classified as type #25, #39, #53, #100, #134). The campaign-based clustering identifies 25 campaigns due to similarities in the website's ownership and registration features. On average, there are 17 websites per campaign cluster, and some significantly larger clusters emerge. The largest campaign (campaign "K") contains 69 websites, all appear to be served their content from within Moldova. All the known domain registrations occur in June, July, and August 2018. The domain names, a factor that was not used to produce the clustering, all explicitly mention "ethereum" or "eth." Fifty-one of the websites are served from the same IP address, and 18 from another similar IP address. All the websites are the advance-fee type #25.

The second-largest campaign (campaign "N") contains 43 websites. This campaign's content appears to be served from IP addresses within Russia. It has domains registered over dates spanning 2018 and 2019. A variety of domain styles exist, including domains targeting Ethereum, Elon Musk, and Tesla. Many of the domains registered in this campaign are associated with the same registrant email address, likely to be a throwaway account. The websites are also all hosted from within the same IP subnet. Different scam types are being run from this campaign, including the *Celebrity/Exchange Giveaway* scam (type #25), the *Exchange Giveaway - Bespoke Format* scam (type #134), and the *Exchange Giveaway - Blog Post* scam (type #100).

Across all the campaigns, it is found that 29% of the identified campaigns are running more than one type of advance-fee scam. This shows that the same entities are running the different types of advance-fee scams and are diversifying their reach. The majority of identified campaigns, 54%, are served solely from IP addresses that appear to be within US. It is found that 25% of the identified campaigns have two or more occurrences of matching Google Analytics IDs.

It should be noted that these different campaigns are likely to represent changing styles of operation by the same entity. We hypothesise that single entities operate multiple of these campaigns; when they are separated into multiple campaigns, it is because there has been a change in operation practices (for example, using a different set of IP addresses or a change of domain registrar).

## C. Analysis of the advance-fee type behaviour

Fig. 3 shows the date of domain registrations and the dates and magnitudes of Bitcoin and Ether inflows (Ether is the name of the main cryptocurrency on the Ethereum blockchain) for the five largest advance-fee types identified. The *Celebrity/Exchange Giveaway* type is active throughout the data period considered, highlighting the continued effectiveness of the type and suggesting that it has operated with impunity. The long lifetime and intensity of domain registrations result in it receiving more incoming transactions than the other types combined. Although it received more transactions, the average USD value of transactions for this type (and the other celebrity-related types) is lower than for the types only targeting exchanges. There was a temporary reduction in the *Celebrity/Exchange Giveaway* domain registrations and cryptocurrency inflows between January 2019 and April 2019. The lull in activity is more pronounced for this type due to the higher frequency of its activity at other times, but is also seen to an extent in the *Celebrity Giveaway – Blog Post* and the *Exchange Giveaway - Blog Post* types. The appearance of the *Exchange Giveaway - Bespoke Format* type late in the dataset highlights it is a more recent threat. Its appearance may also demonstrate that the scams are evolving as this bespoke format appears slightly more complex and convincing than the standard *Celebrity/Exchange Giveaway* type.

The inflows of cryptocurrency occur around (usually after) domain registrations. To better visualise the relationship between domain registrations and particular incoming payments, Fig. 4 shows the number of days between a domain's registration date and incoming payments to cryptocurrency addresses advertised by that domain, for all scam types. It can be seen that 55.4% of incoming cryptocurrency transfers occur within the first week after domain registration, 71.8% of payments occur within the first two weeks and 91.1% of payments occur within the first 30 days. The finding that the majority of payments are received shortly after a domain registration, indicating each domain is active for a short time, aligns with other work which found the lifetime of domains registered for illicit activity is short [22]. The limited lifetime of each domain name also explains the high frequency at which the domains are registered (rather than just one being registered and used throughout the entire duration of a type's life).

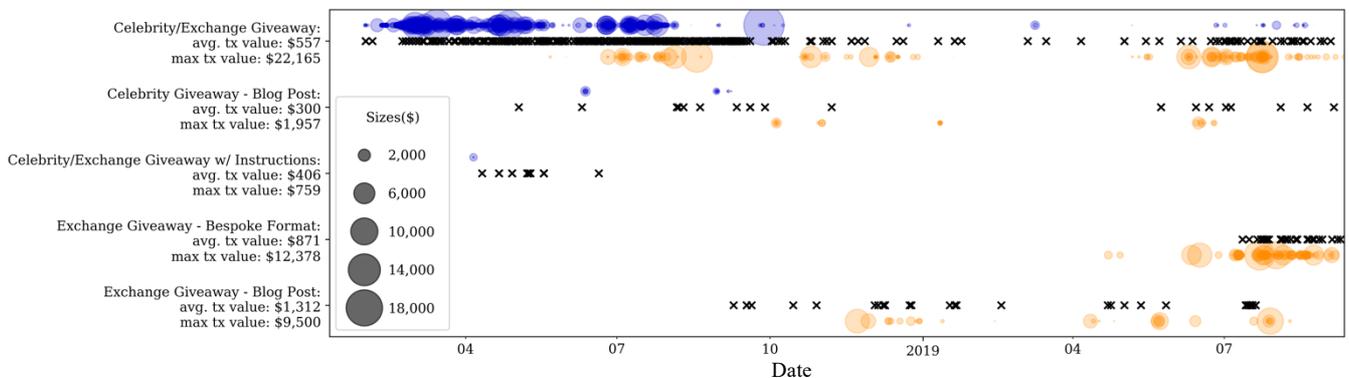

Fig. 3. Domain registrations (marked as "x"), inflows of Ether (blue, above associated domain registrations) and inflows of Bitcoin (orange, below associated domain registrations), over time for the five largest advance-fee types; the size of a bubble represents magnitude in USD. Inflows to addresses associated with multiple scam types are removed to isolate flows to different types.

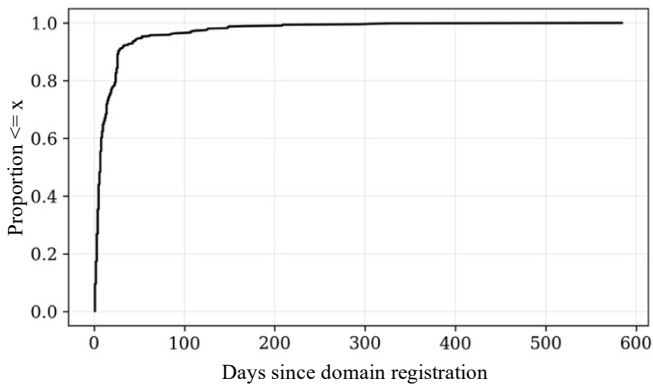

Fig. 4. Emperical Cumulative Distribution Function (ECDF) of the number of days since domain registration that incoming payments to associated cryptocurrency addresses are received on (where available); incoming payments prior to domain registration are excluded.

It should be noted that in some cases addresses receive incoming payments before any recorded domain registrations. There are several possible reasons for this, including that: 1) The address is being used by the scammer before being used for the scam; 2) The address is being used as part of another scam type not considered in this work; 3) The address is being used as part of one of the considered scam types, but not captured in the data collection process. Address reuse between different scam types, including in the case of the *Exchange Giveaway - Bespoke Format* type, is considered later in Section V(D).

At the start of the data period considered in Fig. 3, the majority of inflows are Ether-based (blue-shaded bubbles), whereas over time, the inflows change to being Bitcoin-based (orange-shaded bubbles). It should be noted some Ether inflows still exist throughout the data period, but their magnitudes are smaller than Bitcoin inflows towards the end of the period considered. We hypothesise that the decreasing value of Ether inflows could be a result of the declining Ether price, both in USD and relative to Bitcoin (ETH/BTC), over the time period considered. Generally, the scam websites state the amount to send in units of their target cryptocurrency, and so unchanged units of Ether translate to lower USD values.

The longevity of the *Celebrity/Exchange Giveaway* type provides an opportunity to analyse, for an individual type, how the style of domain registrations changes over time. This provides an insight into the changing tactics of the scams. Near the start of the dataset collected (March 2018 to July 2018), 87% of domain registrations included the word "eth" or "ethereum," whereas towards the end of the dataset collected (May 2019 to September 2019) the percentage is 44%. Over time there is an increase in domain names that: 1) are not specific to one cryptocurrency; 2) relate to particular legitimate cryptocurrency-related events; 3) target celebrities and companies outside of the cryptocurrency ecosystem. The observed proportional reduction in Ethereum-related domain names aligns with the aforementioned reduction in Ether inflows over time. We hypothesise that these scam types transitioned to targeting/receiving Bitcoin rather than Ether potentially due to several factors, including that the scammers aimed to broaden their audience, or more likely, they were following market trends (e.g. the weakening of the ETH/BTC price, mentioned above).

### D. Re-use of blockchain addresses and blockchain clusters

In preparation for the below analysis, addresses that were part of blockchain clusters that are known to be, or exhibit the characteristics of, large cryptocurrency custodial services are removed; such characteristics include abnormally large transaction volumes or received amount. The removal of these clusters is done at this stage because identifying two scam websites attempting to receive payments to two different addresses issued by the same custodial service (for example, trading exchange) does not indicate that the scams are operated by the same entity; if custodial services were kept, the analysis to follow would group unconnected entities together because they have chosen to use the same custodial service (i.e. their addresses fall within the same large blockchain cluster). Under 5% of addresses were removed. Most scams would not operate using addresses hosted by such services as they are at risk of the service seizing their funds. Most of the remaining bitcoin clusters are small; 50% of the clusters contain two or under addresses, 75% of the clusters contain eight or below addresses, and 90% of the clusters contain below twenty-five addresses.

We first consider whether multiple scam types are receiving funds to addresses in the same blockchain cluster. No blockchain cluster reuse between scam types would result in a similarity of 0%. We find that 15.4% of blockchain clusters associated with *Exchange Giveaway - Blog Post* scams are associated with two or more websites advertising such scams; similar, but lower levels of reuse are found for the other scam types. We also find that scams of different types are attempting to receive funds to the same blockchain cluster. For example, *Exchange Giveaway - Blog Post* and the *Exchange Giveaway - Bespoke Format* types share 1.6% of their used blockchain clusters which provides further evidence that the same entities are running different types of the advance-fee scam, similar to the finding that the same IP addresses were running the different types of the scam.

We investigate the previous hypothesis that specific campaigns may appear separate because of different online characteristics (for example, hosting, domain registrar, etc.), but are operated by the same entity. To do this, we consider whether the ownership campaigns identified in Section V(B) use any blockchain addresses which can be linked to the same owner. Fig. 5 shows the individual campaigns as nodes, including campaign "K" and "N" analysed previously, and edges where there is an overlap in certain campaigns' associated blockchain cluster. Campaigns A, B, C, D, G, H, O, and Y form a connected component representing that their addresses are in the same blockchain cluster, meaning the same owner is receiving funds from these campaigns.

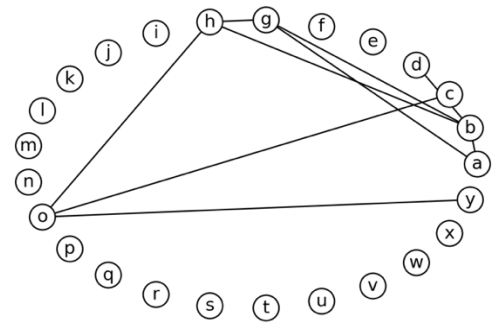

Fig. 5. Blockchain cluster overlap between different campaigns

The ownership information of these campaigns shows several different online infrastructure setups being used by what has now been discovered to be the same controlling entity. This finding shows the same entities are running multiple campaigns using different operational practices and demonstrates the added value of blockchain analysis over merely trying to extract web-based information. Any of the other identified campaigns may be run by the same entity, but their blockchain activity does not provide conclusive evidence.

It is observed that cryptocurrency occasionally moves between addresses and blockchain clusters associated with different scam campaigns. This prompts further investigation into incoming and outgoing flows through exploring the source and destination of funds.

*E. Source of funds*

This section investigates where funds are being sent from before arriving in addresses associated with all the scams collected during this work. To do this, we trace back transactions until they reach a known entity. Table II shows that over 94% of funds received by these scams come from one of the different types of exchanges (either from a fiat-accepting exchange, cryptocurrency exchange, or decentralised exchange).

The next largest source of funds, contributing 2.92%, is from addresses associated with other scams. This initially seems counter-intuitive as the operator of one scam is unlikely to fall for another. Visual inspection of such examples shows blockchain activity manufactured to give the perception of a genuine system. A scammer will send their new scam address X amount and will return Y, in order to appear as if their claims are genuine in an attempt to lure payments in from others.

Fig. 6 shows the geographical location of the exchange's that funds are flowing from, which may provide some indication of the geographical region of victims.[5] Most funds that are received from exchanges come from those based in the U.S., formerly China,[6] and South Korea. We note that funds from U.S. and Chinese exchanges flowed to both Bitcoin and Ether based scams, whereas funds from South Korean exchanges primarily went to Ether based scams.

TABLE II. SOURCE OF FUNDS CATEGORIES

| Category | USD received by scams | % of total USD received by scams |
|---|---|---|
| Fiat-Accepting Exchange | 2,476,100 | 79.32 |
| Cryptocurrency Exchange | 450,300 | 14.43 |
| Scam | 90,900 | 2.91 |
| Miner | 37,800 | 1.21 |
| Wallet Service | 25,500 | 0.82 |
| Payment Service Provider | 11,300 | 0.36 |
| Decentralised Exchange | 9,300 | 0.3 |
| Gambling | 9,200 | 0.29 |
| Ponzi Scheme | 3,100 | 0.1 |
| Dark Market | 2,700 | 0.09 |
| Other | 5,300 | 0.17 |

TABLE III. DESTINATION OF FUNDS CATEGORIES

| Category | USD sent by scams | % of total USD sent by scams |
|---|---|---|
| Fiat-Accepting Exchange | 1,144,900 | 56.68 |
| Gambling | 246,600 | 12.21 |
| Scam | 238,900 | 11.83 |
| Cryptocurrency Exchange | 204,300 | 10.11 |
| Mixer | 54,400 | 2.69 |
| Payment Service Provider | 44,300 | 2.19 |
| Dark Market | 24,900 | 1.23 |
| Phishing | 18,500 | 0.92 |
| Decentralised Exchange | 15,400 | 0.76 |
| Miner | 11,300 | 0.56 |
| Other | 16,500 | 0.82 |

*F. Destination of funds*

Table III shows the destination of funds once leaving addresses associated with the scams. Over 67% of the funds are received by exchanges. Similar preference to send onwards towards exchanges, potentially for immediate cash-out, was seen in previous work while examining illicit flows from dark markets [21]. The destinations can broadly be split into three types of activities: 1) Cashing-out through fiat or by making a purchase, 2) sending funds onwards for another use, or 3) mixing them to obfuscate their final destination. Exchanges and gambling services fall under the first activity where the scammer is likely attempting to exchange their funds for fiat. The act of sending funds onto dark markets and payment service providers indicates that the scammers intend to make illegitimate or legitimate purchases, respectively.

Fig. 6 shows that exchanges based in the U.S. receive the most funds. The majority of funds flowing to U.S.-based exchanges go to two exchanges, potentially because of their lack of blockchain transaction screening. Exchanges in Finland received 17.5% of funds, while they were not a significant source of funds, suggesting those cashing out illicit funds are actively choosing exchanges there. Funds flowing to Finland primarily go to one peer-to-peer fiat-exchange.

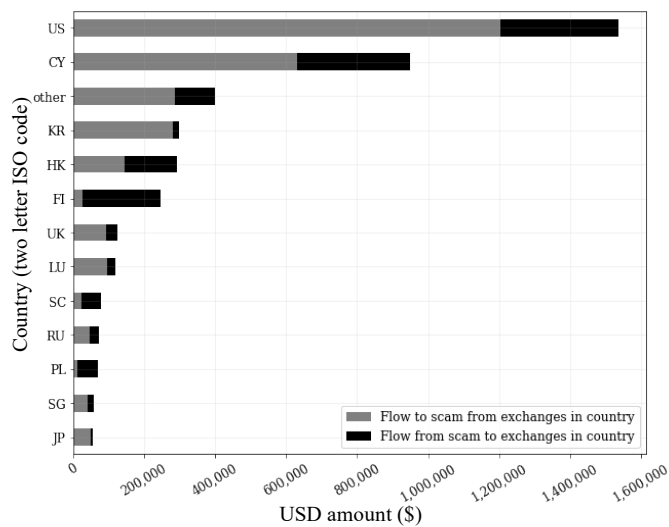

Fig. 6. Country based source and destination of exchange funds

---

[5] It should be noted this does not use the location of victims being scammed, but rather the registered location of the whole exchange. It is only used as an indication, as exchanges usually serve multiple jurisdictions.

[6] Many exchanges originated in China before moving to a range of locations due to preventative regulatory actions within the country. "Formerly China" is used to refer to these exchanges.

## VI. Conclusions

This work provides the first known academic analysis of advance-fee cryptocurrency scams, commonly advertised on social media. Five distinct types of advance-fee scams were found to be used by over 1,000 websites. It was shown that seemingly separate websites are being run under the same organised campaigns, found through clustering based on their hosting and registration features. The features used in the DBSCAN clustering could also be applied to group non-cryptocurrency specific scam websites, however the ability to easily analyse the proceeds raised and easily find further links between campaigns is unique to cryptocurrency-specific scam websites due to the transparent nature of the blockchains used. For example, blockchain analysis linked certain scam websites that appeared initially to be associated with different campaigns to be owned by the same entity (i.e. within the same blockchain cluster) demonstrating that multiple campaigns are run by the same entity using different styles of operation/online infrastructure setups.

Fiat-accepting exchanges are both the largest source of funds and the largest destination of funds. Exchanges may want to be vigilant that scams target their users, but also that scammers attempt to cash out the stolen cryptocurrency through their services. It was also found that scammers are moving cryptocurrency between separate scams, seemingly to generate perceived activity; similar manufactured activity is generated around other illicit blockchain schemes (for example, in smart contract honeypots [14]). The perpetrators utilise the uniquely transparent nature of blockchain transactions to lure victims into believing the scheme is genuine.

At the time of writing, a new type of advance-fee scam has recently come into existence. This new type reuploads videos of legitimate presentations. The sound is sometimes muted, and details of a fictitious giveaway are superimposed over or around the original presentation slides to make it appear that the legitimate speaker is doing a cryptocurrency giveaway; there is the usual requirement that a small amount of cryptocurrency is sent by the victim upfront. This new type of advance-fee scam gives a timely reminder that such scams are still active and that they continue to evolve and advance.

It is hoped the work here will prompt further research into the operation of these low-touch cryptocurrency scams, as further understanding of their operation, including automated identification techniques and prevention, will help for the betterment of the cryptocurrency industry.

## Acknowledgments

We thank Claudio Bellei for his guidance and Alysia Huggins for her technical assistance. We also thank Tom Robinson and Brooke Balza for their feedback.